\documentclass[11pt,draftcls,onecolumn]{IEEEtran}   

\ifCLASSINFOpdf
\else
\fi
\usepackage{amsfonts}
\usepackage{amssymb}
\usepackage{amsmath}
\usepackage{graphicx}
\usepackage{float}
\usepackage{caption}
\usepackage{array}
\usepackage{algorithm}
\usepackage{algorithmic}
\newtheorem{theorem}{Theorem}
\newtheorem{lemma}{Lemma}
\newtheorem{corollary}{Corollary }
\begin{document}

\title{Complex Orthogonal Matching Pursuit and Its Exact Recovery Conditions}
%
%
%
%

\author{Rong Fan, Qun Wan, Yipeng Liu, Hui Chen and Xiao Zhang \\uestc\_fanrong@foxmail.com, wanqun@uestc.edu.cn, yipeng.liu@esat.kuleuven.be, \\huichen\_uestc@yahoo.cn, z-xiao11@mails.tsinghua.edu.cn}

%
%

\markboth{Journal Name}%
{Shell \MakeLowercase{\textit{et al.}}: Bare Demo of IEEEtran.cls for Journals}
%



\maketitle

\textbf{Abstract---}
In this paper, we present new results on using orthogonal matching pursuit (OMP), to solve the sparse approximation problem over redundant dictionaries for complex cases (i.e., complex measurement vector, complex dictionary and complex additive white Gaussian noise (CAWGN)). A sufficient condition that OMP can recover the optimal representation of an exactly sparse signal in the complex cases is proposed both in noiseless and bound Gaussian noise settings. Similar to exact recovery condition (ERC) results in real cases, we extend them to complex case and derivate the corresponding ERC in the paper. It leverages this theory to show that OMP succeed for $k$-sparse signal from a class of complex dictionary. Besides, an application with geometrical theory of diffraction (GTD) model is presented for complex cases. Finally, simulation experiments illustrate the validity of the theoretical analysis.



%
\IEEEpeerreviewmaketitle

\textbf{1. ~Introduction}\par
%
%
%
%
\IEEEPARstart{B}{efore} starting to discuss our problem, we give some symbols illustration. We denote vectors and matrices by boldface lowercase and uppercase letters, respectively. $( \cdot )^{T}$ denotes the transpose operation and $(\cdot)^{H}$ denotes the conjugate transpose operation. Further, $\Vert \cdot \Vert _2$ refers to the $\ell_2$ norm for vectors. $ \mathbf{R} \in \mathbb{R}^{m \times n} $ and $ \mathbf{R} \in \mathbb{C}^{m \times n}$ denote a $m$-by-$n$ real-valued and complex-valued matrix, and  let $\Re\{ \cdot \}$  and $\Im\{ \cdot \}$ be real and imaginary parts, respectively. For a vector $\mathbf{x}$ = $[x_1, x_2, \cdots, x_n]^T \in \mathbb{R}^n$, let \textit{S} = $\lbrace i : \vert x_i \vert \ne 0 \rbrace$ be the support of $ \mathbf{x}$ and let $\mathbf{\Psi}(\textit{S})$ be the set of atoms of $\mathbf{\Psi}$ corresponding to the support \textit{S} and $\mathbf{x}$ is said to be $k$-sparse if the cardinality of the set \textit{S} is no more than $k$ (i.e., $\vert \textit{S} \vert \le k $).\par
Recovery of a high-dimensional sparse signal from a small number of noisy linear measurements, is a fundamental problem in compressive sensing (CS) community. The linear measurement model can be formulated as:
                      \begin{equation}
                        \centering{ \mathbf{y} = \mathbf{\Psi}\mathbf{x} + \mathbf{n} }
                      \end{equation}
where the observation $\mathbf{y} \in \mathbb{R}^m$, the matrix $\mathbf{\Psi}\in \mathbb{R}^{m \times n}$, and the measurement error $\mathbf{n}\in\mathbb{R}^m$. Suppose $\mathbf{\Psi}=[\mathbf{\psi}_1, \mathbf{\psi}_2, \cdots, \mathbf{\psi}_n]$, where $\mathbb{\psi}_i$ denotes the $i$-th column of $\mathbf{\Psi}$. Throughout the paper the matrix  $\mathbf{\Psi}$ and its $i$-th column are called dictionary and the $i$-th atom of $\mathbf{\Psi}$, respectively. The mutual incoherence property (MIP) of dictionary $\mathbf{\Psi}$ is defined as in [1]
\begin{equation}
\centering
\mu (\mathbf{\Psi}) \triangleq \max_{\substack{1 \leqslant i,j \leqslant n \\ i \ne j}}
\frac {\vert \mathbf{{\psi}}_i^T \cdot \mathbf{{\psi}}_j \vert}{\Vert \mathbf{{\psi}}_i \Vert_{2} \cdot \Vert \mathbf{{\psi}}_j \Vert_{2}}
\end{equation}
 CS is to reconstruct the unknown vector $\mathbf{x} \in \mathbb{R}^n$ based on $\mathbf{y}$ and $\mathbf{\Psi}$ . A setting that is of significant interest and challenge is when the dimension $n$ of the signal is much larger than the number of measurements $m$. This problem has received much attention in a number of fields including electrical engineering [2], imaging process [3], statistics and applied mathematics [4], recently.\par
To solve an undetermined system of linear equations in the above form (1), in previous literature, many authors use the OMP algorithm to recover the support of the $k$-sparse signal.
 Compared with other alternative methods (such as [5-8]), a major advantage of the OMP is its low computation complexity. This method has been used for signal recovery and approximation [9-12]. Support recovery has been considered in the noiseless case by Tropp in [10], where it is shown that $\mu <\frac{1}{2k-1}$ is a sufficient condition for recovering a $k$-sparse $\mathbf{x}$ exactly in the noiseless case. Results in [13] imply that this condition is in fact sharp. However, to the author's knowledge, exact recovery condition (ERC) results w.r.t. OMP are derived for real measurement and dictionary.
  When observation $\mathbf{y}$ and dictionary $\mathbf{\Psi}$ as well as noise vector $\mathbf{n}$ are complex, there is no corresponding theory. However, there are many applications in complex settings. Hence, as an extension of the previous theoretical work, we assume that the observation vector $\mathbf{y}$ and dictionary $\mathbf{\Psi}$  are complex. And in the premise we further consider the measurement noise are also complex in the paper. It is the difference between our work and the others and it is also our major contribution in the paper. \par
 According to the above description, with slight abusement of notation, we can directly extend the model (1) to complex value cases as follows.
 \begin{equation}
 \centering
 \mathbf{y} = \mathbf{\Psi}\mathbf{x}+\mathbf{n}
 \end{equation}
 where the observation  $\mathbf{y}$, the matrix $\mathbf{\Psi}$, and the measurement errors $ \mathbf{n}$  are the same dimension as in model (1), respectively. The problem is reformulated into reconstruct the unknown vector $\mathbf{x} \in \mathbb{C}^n$ based on complex vector $\mathbf{y}$ and complex dictionary $\mathbf{\Psi}$.\par
 The paper is organized as follows. In section $\text{2}$, we briefly present the classical OMP algorithm to solve the model (1). We analyze the OMP algorithm ERC for complex value cases in section $\text{3}$. And a geometric theory of diffraction  (GTD)  parametric model is proposed for complex  setting practical application in the section $\text{4}$. Finally, some conclusions and further work are provided in section $\text{5}$.

\textbf{2. ~The OMP Algorithm}\par
 Under the condition (4), the sparse solution can be obtained using OMP algorithm directly. The sparse solution is given by iteratively building up the approximation. The vector $\mathbf{y}$ is approximated by a linear combination of a few atoms in dictionary $\mathbf{\Psi}$ , where the \textit{active set} of atoms is built column by column, in a greedy fashion. At each iteration, a new atom that best correlates with the current residual is added to the \textit{active set}. Here we give a detailed description of the OMP algorithm [14].\par
We assume that the atoms are normalized, i.e., $\Vert \mathbf{\psi}_i \Vert _2 = 1$ , for $i = 1,2,\cdots,n$. We denote the support of $\mathbf{x}$ by $c \subseteq \lbrace 1,2,\cdots,n \rbrace$, which is defined as the set of indices corresponding to the nonzero components of $\mathbf{x}$. $\mathbf{\Psi}(c)$ denotes the matrix formed by picking the atoms of $\mathbf{\Psi}$ corresponding to indices in $c$. In this paper, we use $\mathbf{\psi}_i$ to denote the \textit{i}-th atom of $\mathbf{\Psi}$ in (4). Similarly, we call $\mathbf{\psi}_i$ a correct atom if the corresponding $x_i \ne 0$ and call $\mathbf{\psi}_i$ an incorrect atom otherwise. With slight abuse of notation,
 we use $\mathbf{\Psi}(c)$ to denote both the subset of atoms and the corresponding submatrix of $\mathbf{\Psi}$. The OMP algorithm can be stated as follows in detail (i.e., \textbf{Algorithm 1}). \par

\begin{algorithm}
\caption{: OMP Algorithm}
\label{alg:OMP}
\begin{algorithmic}[1]       
\REQUIRE ~~\\      
The measurement vector $\mathbf{y}$; \\
The dictionary $ \mathbf{\Psi}$ ;\\
the error threshold $\epsilon$; \\
\ENSURE ~~\\     
\STATE Initialize the residual $\mathbf{r}_0 = \mathbf{y}$ and the set of selected atom $\mathbf{\Psi}(c_0)$ = $\phi$. Let iteration counter $i=1$.
\STATE Find the variable $\psi_{t_i}$ that solves the maximization problem
\[ t_i \triangleq \arg\max_{t} \vert \psi_{t}^H \mathbf{r}_{i-1} \vert \]
and add the variable $\psi_{t_i}$ to the set of selected variables. Update $c_i = c_{i-1} \bigcup \lbrace t_i \rbrace$.
\STATE Let $\mathbf{P}_i = \mathbf{\Psi}(c_i)( \mathbf{\Psi}(c_i)^H \mathbf{\Psi}(c_i))^{-1} \mathbf{\Psi}(c_i)^H$
denote the projection onto the linear space spanned by the elements of $\mathbf{\Psi}(c_i)$. Update $\mathbf{r}_i = (\mathbf{I}-\mathbf{P}_i) \mathbf{y}$.
\STATE If the stopping condition is achieved (e.g., $\Vert \mathbf{r}_i \Vert_2 \le \epsilon$), go to 5. Otherwise, set $i=i+1$ and go back to 2 until reaching the given threshold or maximum iterative times.
\STATE Calculate the vector $\mathbf{x}$ with LS method.
\STATE Return $\mathbf{x}$.
\end{algorithmic}
\end{algorithm}


The OMP is a stepwise forward selection algorithm and is easy to implement. A key component of it is the stopping rule which depends on the noise structure. In the noiseless case the natural stopping rule is $\mathbf{r}_i=0$. That is, the algorithm stops whenever $\mathbf{r}_i=0$ is achieved. In this paper, both noiseless case and the case of Gaussian noise with $n_i \sim \textit{CN}(0,\sigma^2)$ are considered. The stopping rule for each case and the properties of the resulting procedure will be discussed in section 3.\par

\textbf{Remark 1:} OMP algorithm starting from $\mathbf{x}=0$. It iteratively constructs a $k$-term approximant by maintaining a set of active atoms (initially empty), and expanding the set by one additional atom at each iteration. The atom chosen at each stage maximally reduces the residual ${\ell}_2$ error in approximating $\mathbf{y}$  from the currently active atoms.
 After constructing an approximant including the new atom, the residual ${\ell}_2$ error is evaluated. If it falls below a specified threshold, the algorithm terminates. It requires $\mathit{O}(nmk)$ flops in total.\par
\textbf{Remark 2}: In fact, one observes that the unknown sparse vector $\mathbf{x}$ is composed of two effective parts which are the support and the non-zero values over the support. Once the support of $\mathbf{x}$ is found via OMP algorithm, the non-zero values of $\mathbf{x}$ are easily determined by least squares (LS) method.\par

\textbf{3 ~Performance Analysis}\par
The performance of the OMP algorithm depends on the probability of selecting a correct atom at each step. The probability is affected by the degree of collinearity among the variables and the noise structure. Ours OMP algorithm analysis will be carried out using the mutual incoherence $\mu(\cdot)$ in (2). Noting that the atoms are normalized and hence it can be rewritten by
\begin{equation}
\centering
\mu (\mathbf{\Psi}) \triangleq \max_{\substack{1 \leqslant i,j \leqslant n \\ i \ne j}}
{\vert \mathbf{{\psi}}_i^H \cdot \mathbf{{\psi}}_j \vert}
\end{equation}
To gain insight on the OMP algorithm and to illustrate the main ideas behind the proofs, it is instructive to provide some technical analysis of the algorithm. The analysis sheds light on how and when the OMP algorithm works properly. However, we must point out that the ERC in noiseless has been verified by Troop in 2004 for real case in [10]. Meanwhile, T. Cai et al. has investigated the properties of the OMP algorithm for bounded noise cases as well as the Gaussian noise case in [13]. In this section, we extend the results to complex case.
 Meanwhile, we also derive the ERC for CAWGN settings. Moreover, it proposes the restrict isometry property (RIP) based bound of the OMP algorithm guaranteeing the exact reconstruction of sparse signals in [14], but it is beyond the scope of our discussion in the paper.\par
\textbf{3.1 ~ERC in the noiseless settings}\par
ERC in noiseless can be posed as a theorem for the success of the OMP as bellow.

\begin{theorem}
For a system of linear equations $\mathbf{y}=\mathbf{\Psi} \mathbf{x} ~(\mathbf{\Psi} \in \mathbb{C}^{m \times n}$  , full-rank with $m < n$), if a solution $\mathbf{x}$ exists obeying
\begin{equation}
\centering
\Vert \mathbf{x} \Vert_0 < \frac{1}{2} \left( 1+ \frac{1}{\mu(\mathbf{\Psi})} \right)
\end{equation}
\end{theorem}

OMP with threshold parameter  $\varepsilon _0 = 0$ is guaranteed to find it exactly, where $\Vert \mathbf{x} \Vert_0$ denotes the non-zero entries in $\mathbf{x}$. We give a proof to Theorem $1$. It is similar to (but not the same as) the Theorem $4.3$ shown in [15]. Here we assume that the dictionary is complex. \par

\begin{proof}
 Without loss of generality, we suppose that the sparsest solution of the linear system is such that all its $k$ non-zero entries are at the beginning of the vector, in decreasing order of the values $\vert x_j \vert$. Thus,
\begin{equation}
\centering
\mathbf{y}=\mathbf{\Psi} \mathbf{x} = \sum_{t=1}^k x_t \mathbf{\psi}_t
\end{equation}
At the first step $(i=0)$ of the algorithm $\mathbf{r}_i = \mathbf{r}_0 = \mathbf{y}$, and the set of computed errors from the sweep step are given by
\begin{equation}
\centering
\epsilon (j)=\min_{z_j} \Vert \mathbf{\psi}_j z_j - \mathbf{y} \Vert _2^2
=\Vert \mathbf{y} \Vert _2^2-(\mathbf{\psi}_j^H \mathbf{y})^2 \ge 0
\end{equation}

To get (8), we utilize the equation $ z_j = \mathbf{\psi}_j^H \mathbf{y}$ and suppose $\Vert \mathbf{\psi}_i \Vert _2^2 =1$, for $i=1,2,\cdots,n$. Thus, for the first step to choose one of the first $k$ entries in the vector (and thus do well),
 we must require that all $i>k$, $\vert \mathbf{\psi}_1^H \mathbf{y} \vert>\vert \mathbf{\psi}_i^H \mathbf{y} \vert$ is satisfied, and substitute it in (7),
 this requirement transforms into
\begin{equation}
\centering
\left\vert \sum_{t=1}^k x_t \mathbf{\psi}_1^H \mathbf{\psi}_t \right\vert
>
\left\vert \sum_{t=1}^k x_t \mathbf{\psi}_i^H \mathbf{\psi}_t \right\vert
\end{equation}

According to (8), we construct a lower bound for the left-hand-side, an upper-bound for the right-hand-side, and then pose the above requirement again. For the left-hand-side we have
\begin{equation}
\centering
\left\vert \sum_{t=1}^k x_t \mathbf{\psi}_1^H \mathbf{\psi}_t \right\vert \ge \vert x_1 \vert \left( 1-\mu(\mathbf{\Psi})(k-1) \right)
\end{equation}
In (9), we exploit triangle inequality theorem and mutual incoherence definition in (4) as well as the decreasing order of the values $\vert x_j \vert$.
Similarly, the right-hand-side term in (8) is bounded by
\begin{equation}
\centering
\left\vert \sum_{t=1}^k x_t \mathbf{\psi}_i^H \mathbf{\psi}_t \right\vert \le k \cdot \vert x_1 \vert \cdot \mu(\mathbf{\Psi})
\end{equation}

For the derivation of (9) and (10), please refer to Appendix A. Using these two bounds plugged into the inequality (8), we obtain
\begin{equation}
\centering
\begin{aligned}
\left\vert \sum_{t=1}^k x_t \mathbf{\psi}_1^H \mathbf{\psi}_t \right\vert &\ge \vert x_1 \vert (1-\mu(\mathbf{\Psi})(k-1))\\
&> \vert x_1 \vert \mu (\mathbf{\Psi})k \ge \left\vert \sum_{t=1}^k x_t \mathbf{\psi}_i^H \mathbf{\psi}_t \right\vert
\end{aligned}
\end{equation}

For the second inequality in (11), we exploit the inequation (5). And then it leads to

\[ 1+ \mu(\mathbf{\Psi})>2\mu(\mathbf{\Psi})k \]

Or equivalently
\begin{equation}
\centering
    k<\frac{1}{2} \left(1+\frac{1}{\mu(\mathbf{\Psi})} \right)
\end{equation}
 which is exactly the condition of sparsity above. This condition guarantees the success of the first stage of the algorithm, which imply that the chosen element must be in the correct support of the sparsest decomposition. \par
\end{proof}

\textbf{3.2 ~ERC in the CAWGN Settings}\par
  Note that the $S=\lbrace i: \vert x_i \vert \ne 0 \rbrace$, and the set of significant or ``correct'' atoms is $\mathbf{\Psi}(S) = \lbrace \mathbf{\psi}_i : i \in S \rbrace$. At each step of the OMP algorithm, the residual vector is projected onto the space spanned by the selected atoms (columns of $\mathbf{\Psi}$).
  Suppose the algorithm selects the correct atoms at the first $t$ steps and the set of all selected atoms at the current step is $\mathbf{\Psi}(c_t)$. Then $\mathbf{\Psi}(c_t)$ contains $t$ atoms and $\mathbf{\Psi}(c_t) \subset \mathbf{\Psi}(S)$. Recall that
\begin{equation}
\centering
\mathbf{P}_t = \mathbf{\Psi}(c_t) \left( \mathbf{\Psi}(c_t)^H \mathbf{\Psi}(c_t) \right)^{-1}\mathbf{\Psi}(c_t)^H
\end{equation}

 is the projection operator onto the linear space spanned by the elements of  $\mathbf{\Psi}(c_t)$. Then the residual after $t$ steps can be written as
\begin{equation}
\begin{aligned}
\Vert \mathbf{r}_t \Vert _2 &= (\mathbf{I}-\mathbf{P}_t)\mathbf{y}\\
&=(\mathbf{I}-\mathbf{P}_t) \mathbf{\Psi} \mathbf{x} + (\mathbf{I}-\mathbf{P}_t) \mathbf{n}\\
&\triangleq \mathbf{s}_t + \mathbf{n}_t
\end{aligned}
\end{equation}
  where $\mathbf{s}_t = (\mathbf{I}-\mathbf{P}_t) \mathbf{\Psi} \mathbf{x}$ is the signal part of the residual and   $\mathbf{n}_t = (\mathbf{I}-\mathbf{P}_t) \mathbf{n}$ is the noise part of the residual.\\
Let
\begin{equation}
\centering
    \alpha_{t,1}=\max_{\mathbf{\psi} \in \mathbf{\Psi}(T)} \lbrace \vert \mathbf{\psi}^H \mathbf{s}_t \vert \rbrace
\end{equation}
\begin{equation}
\centering
    \alpha_{t,2}=\max_{\mathbf{\psi} \in  \mathbf{\Psi} / \mathbf{\Psi}(T)} \lbrace \vert \mathbf{\psi}^H \mathbf{s}_t \vert \rbrace
\end{equation}\\
And
\begin{equation}
\centering
\beta _t = \max_{\mathbf{\psi} \in \mathbf{\Psi}} \lbrace \vert \mathbf{\psi}^H \mathbf{n}_t \vert \rbrace
\end{equation}
It is clear that in order for OMP to select a correct variable at this step, it is necessary to have \par
\begin{equation}
\centering
\max_{\mathbf{\psi} \in \mathbf{\Psi}(S)} \lbrace \vert \mathbf{\psi}^H \mathbf{r}_t \vert \rbrace >
\max_{\mathbf{\psi} \in \mathbf{\Psi}/ \mathbf{\Psi}(S)} \lbrace \vert \mathbf{\psi}^H \mathbf{r}_t \vert \rbrace
\end{equation}

A sufficient condition is $\alpha_{t,1} - \alpha_{t,2} > 2\beta$. This is because $\alpha_{t,1} - \alpha_{t,2} > 2\beta$ implies
\begin{equation}
\centering
\max_{\mathbf{\psi} \in \mathbf{\Psi}(T)} \lbrace \vert \mathbf{\psi}^H \mathbf{r}_t \vert \rbrace \ge \alpha_{t,1} - \beta_t
>\alpha_{t,2} + \beta_t \ge
\max_{\mathbf{\psi} \in \mathbf{\Psi}/ \mathbf{\Psi}(T)} \lbrace \vert \mathbf{\psi}^H \mathbf{r}_t \vert \rbrace
\end{equation}

However, exploiting Lemma 4 and Lemma 5 results in [13], we have the following results:\par
\begin{lemma}
 The minimum eigenvalue of $\mathbf{\Psi}(S)^H \mathbf{\Psi}(S)$ is less than or equal to the minimum eigenvalue of $\mathbf{\Psi}(u_t)^H (\mathbf{I}-\mathbf{P}_t) \mathbf{\Psi}(u_t)$. And a sufficient condition for selecting a correct atom at the current step is
 $\Vert \mathbf{x}(u_t) \Vert_2 > \frac{2\sqrt{k-t} \beta_t}{1-(2k-1)\mu(\mathbf{\Psi})}$.
 $\mathbf{\Psi}(u_t) \triangleq \mathbf{\Psi}(S)/ \mathbf{\Psi}(c_t)$ denote the set of significant atoms that are yet to be selected and $\mathbf{x}(u_t)$ denotes the corresponding linear coefficients.
 \end{lemma}

 The complex Gaussian noise case is of particular interest in this paper. To simplify deviation, we present an important result on bound noise cases given in [13].
\begin{lemma}
Suppose   $\Vert \mathbf{n} \Vert _2 \le b_2$ and $\mu(\mathbf{\Psi})<\frac{1}{2k-1}$. Then the OMP algorithm with the stopping rule  $\Vert \mathbf{r}_i \Vert _2 \le b_2$ recovers exactly the true subset of correct atoms  $\mathbf{\Psi}(S)$ if all the nonzero coefficients $x_i$ satisfy
$\vert x_i \vert > \frac{2b_2}{ \left( 1-(2k-1) \mu( \mathbf{\Psi}) \right)}$.
\end{lemma} \par
 The results in Lemma 2 can be applied to the case where noise is Gaussian. This is due to the fact that Gaussian noise is ``essentially bounded" as it proved in [16]. Although Lemma 2 is derived for the real cases in [13],
  it also holds in complex AWGN cases. The proof is in Appendix B. Suppose the noise vector follows complex Gaussian distribution, i.e., $\mathbf{n} \sim \mathit{CN} (0,\sigma^2 \mathbf{I}_m)$ and each $n_i$ is i.i.d. Define the following bounded set

\begin{equation}
\centering
\textbf{B}_1 = \left\lbrace \mathbf{n} : \Vert \mathbf{n} \Vert_2 \le \sigma \sqrt{ \left( m+ \sqrt{2m \cdot \ln(2m)} \right)} \right\rbrace
\end{equation}
 Then we have the following result.

\begin{theorem}
Suppose noise vector in (4) $\mathbf{n} \sim \mathit{CN} (0,\sigma^2 \mathbf{I}_m)$, entries of noise are \textit{i.i.d}, and real part as well as imaginary part in $n_k$ are also \textit{i.i.d}. Then the Gaussian error satisfies
\begin{equation}
\centering
\mathbf{P} (\mathbf{n} \in \mathbf{B}_1) \ge 1- \frac{1}{2\sqrt{\pi \cdot \ln(2m)}}
\end{equation}
\end{theorem}
    The proof is in Appendix C. \par
    Let the bound noise be a different form, then it could directly get a different result in Corollary 1.\par

   \begin{corollary}
   Suppose noise vector in (4) $\mathbf{n} \sim \mathit{CN}(0,\sigma^2 \mathbf{I}_m)$, entries of noise are \textit{i.i.d}., real part as well as imaginary part in $\mathbf{n}_k$ are also \textit{i.i.d}, and if
    $\mathbf{B}_2 = \left\{ \mathbf{n}: \Vert \mathbf{n} \Vert_2 \le \sigma \sqrt{m+\frac{1}{2} \cdot \sqrt{m \cdot \ln(m)}} \right\}$. Then the Gaussian error satisfies
\begin{equation}
\mathbf{P}(\mathbf{n} \in \mathbf{B}_2) \ge 1- \sqrt{\frac{2}{\pi \cdot \ln(m)}}
\end{equation}
\end{corollary}
 The proof is in Appendix D.\par
 \textit{Lemma 2} suggests that one can apply the results obtained for the bounded error case to solve the complex Gaussian noise problem. We directly apply the results for $\mathit{l}_2$ bounded noise case (Lemma 2) and Theorem 2 to get the ERC in CAWGN cases.\par
\begin{theorem}
Suppose $\mathbf{n} \sim \mathit{CN} (0,\sigma^2 \mathbf{I}_m)$, $\mu(\mathbf{\Psi})<\frac{1}{2k-1}$, and all the nonzero coefficients $x_i$ satisfy
\begin{equation}
\centering
\vert x_i \vert \ge \frac{2\sigma \sqrt{\left( m+\sqrt{2m \cdot \ln(2m)} \right)}} {1-(2k-1)\mu(\mathbf{\Psi})}
\end{equation}
Then OMP algorithm with the stopping rule $\Vert \mathbf{r}_i \Vert \le \sigma \sqrt{ \left( m+ \sqrt{2m \cdot \ln(2m)} \right)}$ can select the true subset $\mathbf{\Psi}(\mathit{S})$ with probability at least $1-\frac{1}{2\sqrt{\pi \ln(2m)}}$ .
\end{theorem} \par
Meanwhile, with the results in Lemma 2 and Corollary 1 we can obtain a different ERC in CAWGN cases.\par

\begin{theorem}
Suppose  $\mathbf{n} \sim \mathit{CN} (0,\sigma^2 \mathbf{I}_m)$, $\mu(\mathbf{\Psi})<\frac{1}{2k-1}$ and all the nonzero coefficients $x_i$ satisfy
\begin{equation}
\centering
\vert x_i \vert \ge \frac{2\sigma \sqrt{ m+\frac{1}{2} \cdot \sqrt{m \cdot \ln(m)}}} {1-(2k-1)\mu(\mathbf{\Psi})}
\end{equation}
Then OMP algorithm with the stopping rule $\Vert \mathbf{r}_i \Vert \le \sigma \sqrt{ m+ \frac{1}{2} \cdot \sqrt{m \cdot \ln(m)}}$ selects the true subset   $\mathbf{\Psi}(\mathit{S})$ with probability at least $1-\sqrt{\frac{2}{\pi \ln(m)}}$.\par
\end{theorem}
\par We omit the proof Theorem 3 and Theorem 4 because it is obvious and easy.\par
However, before we end the theoretical analysis, we should mention that all the results derived in this paper are worst-case ones, implying that the kind of guarantees we obtain are over-pessimistic, as they are supposed to hold for all signals and for all possible supports of a given cardinality.
 Besides, compared with the ERC in [13], the derived ERC recovery success probability is larger. The mainly reason is due to in complex cases, the measurement vector, dictionary, the high dimension sparse unknown vector as well as noise vector are assumed complex. If all of them is real, the ERC also reduces to those results in [13].
 However, the dictionary MIP is a more fundamental role and the constraint relationship $\mu (\mathbf{\Psi}) < \frac{1}{2k-1}$ is unchanged.

\textbf{4. ~An Application for Complex Case}\par
In this section, we present an exact application of the OMP algorithm for complex case via GTD model which is widely used by radar imaging community [17]. The GTD model is proposed in the literature [18] and [19]. We give the mathematical description about the model in radar imaging, firstly.

\textbf{4.1 ~Simulation Application Formulation}\par
In the paper, ideal point scattering mechanism is considered. It assumes that the measured scattering data from $d$ scattering centers at $M$ sampled frequency points $f_m ~(m = 0,1,\cdots,M-1)$ and one aspect angle are given by [20]\par
\begin{equation}
\centering
y_m =
 \sum_{p=1}^d A_p \cdot \exp \left\lbrace -j \frac{4\pi}{s} f_m r_p \right\rbrace
\end{equation}

The model parameters $\lbrace A_p, r_p \rbrace _{p=1}^d$ characterize the $d$ individual scattering centers intensity and the distance from reference center on the target to scatterers, respectively. $A_p$ is a complex scalar providing the magnitude. $f_m$ is the $m$-th measurement frequency. $s$ is the speed of light in free space. Using equation $\tau_p = \frac{2r_p}{s}$, the model (25) can be formulated as the compact matrix form in noise setting,

\begin{equation}
\centering
\mathbf{y} = \mathbf{\Psi x}+\mathbf{n}
\end{equation}

where $\mathbf{y} \in \mathbb{C}^{m \times 1}$ is the observation vector in frequency; $\mathbf{\Psi} \in \mathbb{C}^{m \times n}$ is the transform matrix with the $l$-th row and $p$-th column element is
\begin{equation}
\centering
[\mathbf{\Psi}]_{l,p} = \exp \{-j2\pi f_l \tau_p\}
\end{equation}

$\mathbf{x} \in \mathbb{C}^{n \times 1}$ corresponds to magnitude of the scattering centerer.  $\mathbf{u}\in \mathbb{C}^{m \times 1}$ is stochastic measurement noise vector; assuming that $\mathbf{n} \sim \mathit{CN}(0,\sigma^2\mathbf{I}_m)$ is a vector of \textit{i.i.d} random variables.
Note that all the columns are normalized (i.e., $\Vert \mathbf{\psi}_i \Vert = 1$ for $i =1,2,\cdots,n $), and the measurement errors $\mathbf{n} \in \mathbb{C}^{m \times 1}$. In (26), obviously, it is a problem to recover a high-dimensional sparse signal based on a small number of linear measurements, in noise settings.\par
\textbf{4.2 ~Simulation Results}\par

\begin{figure}[!t]
\centering
\includegraphics[width=3.5in]{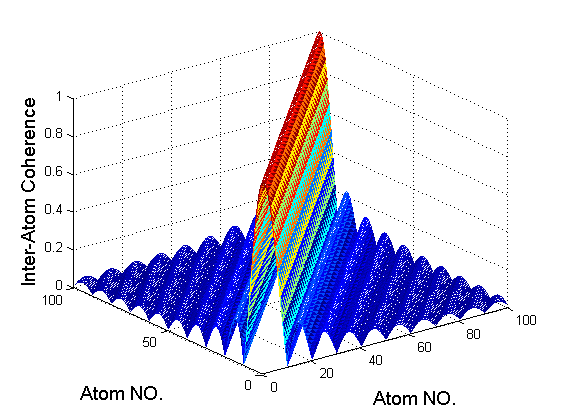}\vspace{0.0mm}
\caption{inter-atom mutual inference property (3D)}
\label{fig_sim}\vspace{0.0mm}
\end{figure}

\begin{figure}[!t]
\centering
\includegraphics[width=3.5in]{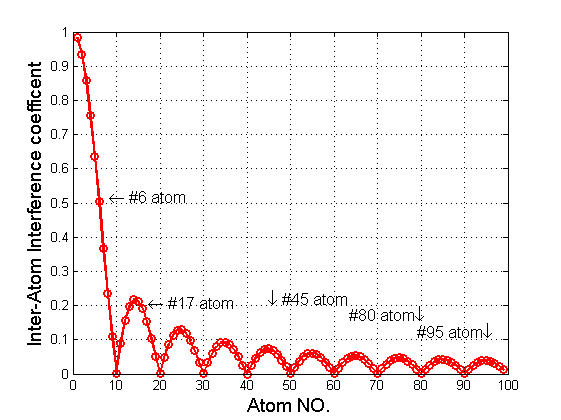}\vspace{0.0mm}
\caption{inter-atom mutual inference property (2D)}
\label{fig_sim}\vspace{0.0mm}
\end{figure}

\begin{figure}[!t]
\centering
\includegraphics[width=3.5in]{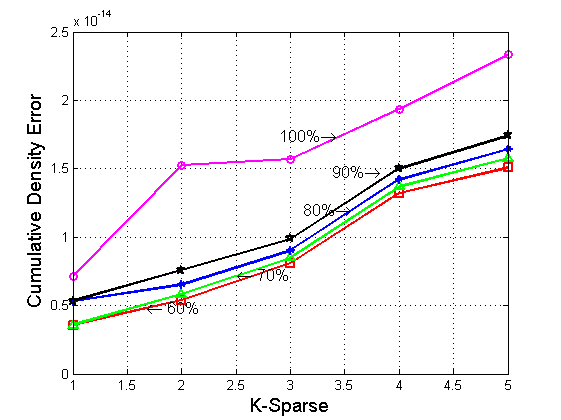}\vspace{0.0mm}
\caption{CDE vs. different $k$-sparse (noiseless)}
\label{fig_sim}\vspace{0.0mm}
\end{figure}

\begin{figure}[!t]
\centering
\includegraphics[width=3.5in]{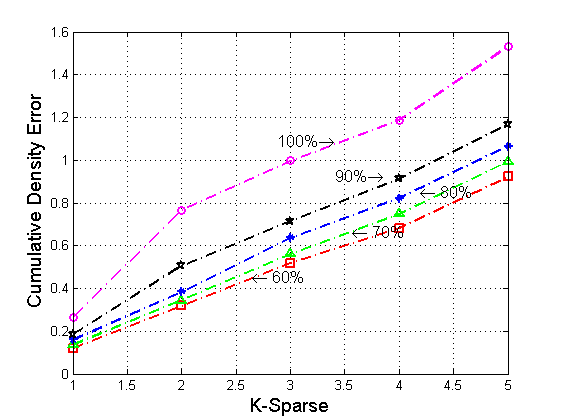}\vspace{0.0mm}
\caption{CDE vs. different $k$-sparse (noiseless)($SNR=20$dB)}
\label{fig_sim}\vspace{0.0mm}
\end{figure}

\begin{figure}[!t]
\centering
\includegraphics[width=3.5in]{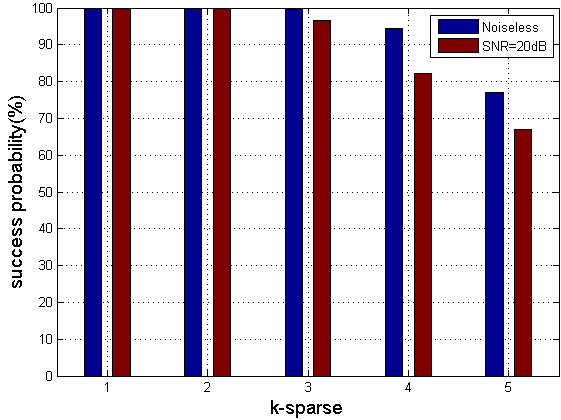}\vspace{0.0mm}
\caption{success recover probability w.r.t different $k$-sparse}
\label{fig_sim}\vspace{0.0mm}
\end{figure}

In this subsection, 10,000 trails Monte Carlo simulation has been done for confirming the previous theoretical analysis in section 3 via an exact application introduced in subsection 4.1. In the simulation, the measured frequency band ranges from 1GHz to 1.3GHz in $L$ band, where the start frequency is $f_0 =1$GHz and frequency sampling interval is $10$MHz.
 Then 30 complex frequency samples can be measured. Furthermore, we assume the target is 5m length and composed of one to five scatter points located at 0.3m, 0.85m, 2.25m, 4.0m and 4.75m to target front-end respectively. The measured samples in frequency are contaminated by CAWGN with $SNR=20dB$ and noiseless, respectively. \par
The dictionary mutual incoherence coefficient shown in Fig. 1 and Fig. 2 are calculated by formula (4). Fig.1 shows three-dimension plot among all atoms of $\mathbf{\Psi}$. Fig. 2 is two-dimension situation respectively. We can see that when the interval between two atoms is smaller, the coherence is larger. Besides, once determining the support of sparse vector $\mathbf{x}$ with OMP,
 we further calculate the non-zero values over this support of $\mathbf{x}$ with LS method. Hence, Fig. 3 and Fig. 4 also present the cumulative distribute error (CDE) w.r.t different $k$-sparse settings (i.e., $k$ varies from 1 to 5 in simulation) both in noiseless and $SNR=20dB$, respectively.
  Finally, Fig. 5 presents the simulation results on recovery probability w.r.t different $k$-sparse settings both in noiseless and $SNR=20dB$. As Fig. 2 implies, we give some remarks about OMP algorithm as bellow. \par
\textbf{Remark 4:} Fig. 3 and Fig. 4 show the CDE are a monotonic increasing relative to sparse degree $k$. It is easy to understand, because with sparse degree $k$ increases from 1 to 5, it is harder and harder to make $\mu(\mathbf{\Psi})$ satisfy with inequality constraint (12). In Fig. 3, we consider noiseless case, while Fig. 4 presents CDE for the case $SNR = 20dB$.\par

\textbf{Remark 5:} If only one scatter point is in $\mathbf{x}$, it is to recover a $1$-sparse vector. Obviously, the recover success probability must be 100\% in this setting, because $\mathbf{x}$ is a $1$-sparse,
 dictionary mutual incoherence among all atoms $ \mu(\mathbf{\Psi})<1$ always holds. Even in the noise settings with $SNR=20dB$, hence it can select correct support of $\mathbf{x}$ via the OMP algorithm. Meanwhile, when $\mathbf{x}$ is $2$-sparse,
  it is also recovered with the probability 100\% no matter in noiseless or $SNR=20dB$ settings. As a matter of fact, it can be recover because inter-atom incoherence $\mu(\mathbf{\Psi})<1/3$ is always satisfied only if the interval in support is large enough so that the two selected atoms mutual incoherence satisfy with  $\mu(\mathbf{\Psi}) <1/3 $.
  Obviously, the incoherence condition is satisfied (see from Fig. 2).\par

\textbf{5. ~Conclusion and Future Work}\par
In this paper, some new results on using OMP algorithm to solve the sparse approximation problem over redundant dictionaries with complex cases are presented. With the mutual incoherence property to quantify inner-atom interference (IAI) level in dictionary,
it provides a sufficient condition under which OMP can recover the optimal representation of an exactly sparse signal in the complex settings. It leverages this theory that OMP can succeed
for $k$-sparse signal from a class of dictionary with high probability. More importantly, the new proposed ERC in complex cases completes the existed ERC of OMP. It makes OMP ERC become more complete. In the end, we confirm the correction of theoretical analysis via simulation experiments.\par
Some future work will be addressed. First, we only consider ERC of classical OMP algorithm in complex cases and not considering IAI yet (see from Fig. 1 and Fig. 2). In fact, if interval of a two non-zero elements in $\mathbf{x}$ is small even the two elements is adjacent each other,
it cannot recover with high probability, which is mainly caused by IAI. Although there are much literatures about imitate IAI with sensing dictionary such as [21] and [22], all of them are for real dictionary. Similarly, how to imitate IAI in complex case is worthy of research.
 Second, we just assume that there is only one scatter type, but in fact there are a few of scatter types such as [19] provided. If all scatter types are considered, the dictionary has a large scale. Hence, in this situation, how to deduce dictionary dimension (in other words, how to reduce computation cost) is also a problem.
 Third, we use LS method to recover non-zeros values in sparse vector $\mathbf{x}$. How to reduce recover error of non-zeros value is also a question to consider in noise settings. Besides, OMP works correctly for a fixed signal and measurement matrix with high probability, and so it must fail for some sparse signals and matrices [23].
  While in complex settings it is also having this problem indeed, as far as our known.

\section*{Acknowledgment}
The authors would like to thank the anonymous reviews for their comments that help to improve the quality of the paper.
This research was supported by the National Natural Science Foundation of China (NSFC) under Grant 61172140, and `985' key projects
for excellent teaching team supporting (postgraduate) under Grant A1098522-02. Yipeng Liu is supported by FWO PhD/postdoc grant: G.0108.11 (Compressed Sensing).

\appendices
\section{Proof of (10) and (11)}
In (9), for the left-hand-side we have
  \begin{equation}
\begin{aligned}
\left\vert \sum_{t=1}^k x_t \mathbf{\psi}_1^H \mathbf{\psi}_t \right\vert
& > \vert x_1 \vert - \sum_{t=2}^k \vert x_t \vert \vert \mathbf{\psi}_1^H \mathbf{\psi}_t \vert \\
& \ge \vert x_1 \vert - \sum_{t=2}^k \vert x_t \vert \mu(\mathbf{\Psi}) \\
& \ge \vert x_1 \vert \left(1-\mu(\mathbf{\Psi})(k-1) \right)
\end{aligned}
  \end{equation}

Here we have exploited the definition of the mutual-coherence (4), and the descending ordering of the values $\vert x_j \vert$. Similarly, In (10), the left-hand-side term is bounded by
  \begin{equation}
\begin{aligned}
\left\vert \sum_{t=1}^k x_t \mathbf{\psi}_1^H \mathbf{\psi}_t \right\vert
& \le \sum_{t=1}^k \vert x_t \vert \vert \mathbf{\psi}_i^H \mathbf{\psi}_t \vert \\
& \le \sum_{t=1}^k \vert x_t \vert \mu(\mathbf{\Psi}) \\
& < \vert x_1 \vert \mu(\mathbf{\Psi})k
\end{aligned}
  \end{equation}
\section{Proof of Lemma 2}
 It follows from the assumption $\Vert \mathbf{n} \Vert _2 \le b_2$ that
\[ \Vert \mathbf{n}_t \Vert_2 \le \Vert (\mathbf{I}-\mathbf{P}_t) \mathbf{n} \Vert_2 \le \Vert \mathbf{n} \Vert_2 \le b_2\]

Let $\mathbf{\psi}_i$ be any column of $\mathbf{\Psi}$. Then,
\[ \vert \mathbf{\psi}_i^H \mathbf{n}_t \vert
\le
\Vert \mathbf{\psi}_i \Vert_2 \Vert \mathbf{n}_t \Vert_2 \le b_2\]

This means $\beta_t \le b_2$. It follows from Lemma 1 that for any $t<k$. $\Vert \mathbf{x}(u_t) \Vert_2 > \frac{2\sqrt{k-t}\beta_t}{1-(2k-1)\mu(\mathbf{\Psi})}$ implies that a correct atom will be selected at this step. So $\vert x_i \vert_2 \ge \frac{2b_2}{1-(2k-1)\mu(\mathbf{\Psi})}$ for all nonzero coefficients $x_i$ ensures that all the $k$ correct atoms will be selected in the first $k$ steps.\par

Let us now turn to the stopping rule. Let $\mathbf{P}_k$ denote the projection onto the linear space spanned by $\mathbf{\Psi}(\mathit{T})$. Then $\Vert (\mathbf{S}- \mathbf{P}_k) \mathbf{n}\Vert_2 \le \Vert \mathbf{n} \Vert_2 \le b_2$. When all the $k$ correct atoms are selected, the $l_2$ norm of the residual will be less than $b_2$. Hence the algorithm stops. It remains to be shown that the OMP algorithm does not stop early.\par

Suppose the algorithm has run $t$ steps for some $t<k$. We will verify that $\Vert \mathbf{r}_t \Vert_2 > b_2$. So OMP does not stop at the current step. Again, let $\mathbf{\Psi}(u_t)$ denote the set of unselected but correct variable and $\mathbf{x}(u_t)$ denote the corresponding coefficients. Note that
\begin{equation}
\begin{aligned}
\Vert \mathbf{r}_t \Vert_2
& = \Vert (\mathbf{I} - \mathbf{P}_t) \mathbf{\Psi}\mathbf{x} + (\mathbf{I} - \mathbf{P}_t) \mathbf{n}_2\Vert_2\\
& \ge \Vert (\mathbf{I} - \mathbf{P}_t) \mathbf{\Psi}\mathbf{x} \Vert_2 - \Vert (\mathbf{I} - \mathbf{P}_t) \mathbf{n} \Vert_2\\
& \ge \Vert (\mathbf{I} - \mathbf{P}_t) \mathbf{\Psi}(u_t) \mathbf{x}(u_t) \Vert_2 - b_2
\end{aligned}
\end{equation}
It follows from Lemma 1 that
\begin{equation}
\begin{aligned}
\Vert (\mathbf{I} - \mathbf{P}_t) \mathbf{\Psi}(u_t) \mathbf{x}(u_t) \Vert_2 & \ge \lambda_{min} \Vert \mathbf{x}(u_t)\Vert_2\\
&\ge \left( 1-(k-1)\mu(\mathbf{\Psi})\right) \frac{2b_2}{\left( 1-2(k-1)\mu(\mathbf{\Psi})\right)}\\
& \ge 2b_2
\end{aligned}
\end{equation}
So,
\[ \Vert \mathbf{r}_t \Vert_2 \ge \Vert (\mathbf{I}-\mathbf{P}_t) \mathbf{\Psi}(u_t) \mathbf{x}(u_t)\Vert_2 -b_2 > b_2\]
and the lemma is proved.

\section{Proof of Theorem 2}
Without loss of generality, let $k$-th element in noise vector $\mathbf{n}$ be $\mathbf{n}_k$. Then $\mathbf{n}_k$ is independent identical distribute and $\mathbf{n}_k \sim \mathit{CN}(0,\sigma^2)$.\par
As real part and imaginary part in $\mathbf{n}_k$ are also \textit{i.i.d}, then we have,
\[ \Re \{ \mathbf{n}_k\} \sim N \left(0,\frac{\sigma^2}{2} \right)\]

and

\[ \Im \{ \mathbf{n}_k\} \sim N \left(0,\frac{\sigma^2}{2} \right)\]

Equivalently, we further have,
\[ \frac{\sqrt{2}}{\sigma}\cdot \Re \{ \mathbf{n}_k\} \sim N(0,1)\]
and
\[ \frac{\sqrt{2}}{\sigma}\cdot \Im \{ \mathbf{n}_k\} \sim N(0,1)\]

Considering the independence between real part and imaginary part in $\mathbf{n}$, we have the following equations，
\[\Vert \mathbf{n} \Vert_2^2 = \Vert \Re \{ \mathbf{n} \}\Vert_2^2 +\Vert \Im \{ \mathbf{n} \}\Vert_2^2 \]

Then, we can further get,
\begin{equation}
\begin{aligned}
Y_{2m}
& \triangleq \frac{2}{\sigma^2} \Vert \mathbf{n}\Vert_2^2 \\
& = \frac{2}{\sigma^2} \left( \sum_{k=1}^m \Re\{ \mathbf{n}_k\}^2 + \sum_{k=1}^m \Im \{ \mathbf{n}_k\}^2 \right)\\
&= \left( \sum_{k=1}^m \left( \frac{\sqrt{2}}{\sigma} \cdot \Re \{ \mathbf{n}_k\}\right) +
\sum_{k=1}^m \left( \frac{\sqrt{2}}{\sigma} \cdot \Im \{ \mathbf{n}_k\}\right) \right)
\end{aligned}
\end{equation}
Obviously,
\[ Y_{2m} \sim \chi^2(2m)\]

It follows from Lemma 4 in [24] that for any $\lambda >0$  and using the technical inequality relationship $\ln(1+\lambda)< \lambda$, it holds when  $\lambda >-1$ ~but $\lambda \ne 0$ ，we can get,
\begin{equation}
\begin{aligned}
&\mathit{P} \left( Y_{2m} >(1+\lambda)2m \right)\\
& \le \frac{1}{\sqrt{\pi}} \cdot \frac{1}{\lambda} \cdot \frac{1}{\sqrt{2m}} \left( (1+\lambda)e^{-\lambda}\right)^m \\
& = \frac{1}{\lambda \sqrt{2\pi m}}  \exp \left\{ \ln \left( (1+\lambda) e^{-\lambda}\right)^m \right\} \\
& = \frac{1}{\lambda \sqrt{2\pi m}}  \exp \left\{ -m \left( \lambda - \ln(1+\lambda) \right) \right\} \\
& \le \frac{1}{\lambda \sqrt{2\pi m}}
\end{aligned}
\end{equation}

~~~~Hence,
\begin{equation}
\begin{aligned}
& \mathit{P} \left( \Vert \mathbf{n} \Vert_2^2  \le \frac{\sigma^2}{2} \left( 2m+2\sqrt{2m \cdot \ln(2m)}\right) \right)\\
& = \mathit{P} \left( Y_{2m} \le \left( 2m+2\sqrt{2m \cdot \ln(2m)}\right) \right) \\
& = 1- \mathit{P} \left( Y_{2m} > \left( 2m+2\sqrt{2m \cdot \ln(2m)}\right) \right) \\
& = 1- \mathit{P} \left( Y_{2m} > 2m \left( 1+ \sqrt{2m^{-1} \cdot \ln(2m)}\right) \right) \\
& = 1- \mathit{P} \left( Y_{2m} > 2m(1+\lambda) \right) \\
& \ge 1- \frac{1}{\lambda \sqrt{2\pi m}}
\end{aligned}
\end{equation}

Finally, we substitute $\lambda = \sqrt{2m^{-1} \cdot \ln(2m)}$ into the inequality above. It becomes
\[ \mathit{P} \left( \Vert \mathbf{n} \Vert_2 \le \sigma \sqrt{\left( m+ \sqrt{2m \cdot \ln(2m)} \right)} \right) > 1- \frac{1}{2\sqrt{\pi \ln(2m)}}\]

Hence, the results of theorem 1 holds.

\section{Proof of Corollary 1}

For the corollary, similar to the proof of the theorem 2, the procedure is as follows,
\begin{equation}
\begin{aligned}
& \mathit{P} \left( \Vert \mathbf{n} \Vert_2^2  \le \frac{\sigma^2}{2} \left( 2m+ \sqrt{m \cdot \ln(m)}\right) \right)\\
& = \mathit{P} \left( Y_{2m} \le \left( 2m+ \sqrt{m \cdot \ln(m)}\right) \right) \\
& = 1- \mathit{P} \left( Y_{2m} > \left( 2m + \sqrt{m \cdot \ln(m)}\right) \right) \\
& = 1- \mathit{P} \left( Y_{2m} > 2m \left( 1+ \frac{1}{2} \cdot \sqrt{m^{-1} \cdot \ln(m)}\right) \right) \\
& = 1- \mathit{P} \left( Y_{2m} > 2m(1+\lambda) \right) \\
& \ge 1- \frac{1}{\lambda \sqrt{2\pi m}}
\end{aligned}
\end{equation}

Similarly, substitute $\lambda = \frac{1}{2}\sqrt{m^{-1} \cdot \ln(m)}$ into the inequality above, we can get the results:
\[ \mathit{P} \left( \Vert \mathbf{n} \Vert_2 \le \sigma \sqrt{\left( m + \sqrt{2m \cdot \ln(2m)} \right)} \right) > 1- \frac{2}{\sqrt{\pi \ln(m)}}\]

\ifCLASSOPTIONcaptionsoff
  \newpage
\fi

\end{document}